# Contact Instability in Adhesion and Debonding of Thin Elastic Films

Manoj Gonuguntla[1], Ashutosh Sharma[1],[*] Jayati Sarkar[1], Subash A. Subramanian[1], Moniraj Ghosh[1] and Vijay Shenoy[2]

[1] Department of Chemical Engineering, Indian Institute of Technology Kanpur, UP 208016, India

[2] Material Research Centre, Indian Institute of Science, Bangalore 560 012, India

Based on experiments and 3-D simulations, we show that a soft elastic film during adhesion and debonding from a rigid flat surface undergoes morphological transitions to pillars, labyrinths and cavities, all of which have the same lateral pattern length scale, $\lambda$ close to $\lambda/H \sim 3$ for thick films, $H > 1$ μm. The linear stability analysis and experiments show a new thin film regime where $\lambda/H \approx 3 + 2\pi (\gamma/3\mu H)^{1/4}$ ($\gamma$ is surface tension, $\mu$ is shear modulus) because of significant surface energy penalty (for example, $\lambda/H \approx 6$ for $H = 200$ nm; $\mu = 1$MPa).

PACS numbers: 82.35.Gh, 82.45.Mp, 68.35.-p, 05.65.+b

Self-organized structures in thin soft confined films are of scientific and technological importance in variety of settings such as wetting [1], adhesion [2,3], friction, patterning [4-7], sensors, lab-on-a-chip devices, microfluidics, MEMS, etc. Thin soft films confined by the long-range forces such as the van der Waals interactions and electric fields are inherently metastable or unstable and thus readily self-organize their shapes to reduce the total energy. Of particular interest here is the surface of a soft elastic solid film, which when brought in contact with a flat rigid surface, forms a short-wave isotropic structure [2,3,8,9]. All of the previous experimental studies [2] of flat elastic contact are however limited to the observation of relatively thick (> 10 µm) and closely adhering films where a labyrinth structure or fingers are reported. Theoretical studies based on the linear stability analysis [3,8] and 2-D simulations [9] reveal that such instabilities are a result of a competition between the attractive adhesive force and the restraining elasticity of the film and the effect of surface tension is not important. We investigate, both experimentally and theoretically, the morphology and length scales of the self-organized structures at a soft elastic interface during the entire cycle of adhesion and debonding with a flat stamp. Three distinct morphological phases (pillars, labyrinths and cavities) are uncovered in adhesion-debonding cycle, all of which have about the same mean periodicity, $\lambda$. We resolve the invariance of length scales and the physics of observed morphological transitions during the process of adhesion and debonding by energy minimizing 3-D nonlinear simulations. Further, we show that the property independent short-wave scaling implied by the linear stability analysis [8] for the lateral length scale ($\lambda \sim 3\,H$) changes completely to a long-wave scaling with nonlinear dependence on film thickness, surface tension and shear modulus for sub-micron films because of a significant surface energy contribution. The above issues of morphological transitions and the lengthscales are also of utmost importance in developing a potential sub-micron patterning technique based on the alignment of the elastic contact instability by a pre-patterned stamp.

It may be noted that the mechanisms and morphologies of instability in a confined elastic solid film [2,3,8,9] are entirely different from a thin liquid viscous film [1,4,7]. Liquid films destabilize by a long-wave instability [1], which is very sensitive to the exact nature of the confining-field and the surface tension [1,4,7]. We show here that the strength of the confining field has no effect on the pattern wavelength in a soft elastic film regardless of its thickness, but the surface energy for sub-micron films changes the wavelength and its scaling very profoundly. Interestingly, in contrast to extensive work on self-organization in thin liquid films, there are very few known mechanisms [2,6] for the generation of self-organized structures in thin elastic solid films and no understanding of the role of surface tension in this phenomenon.

Filtered Sylgard-184 (a two part PDMS elastomer; Dow Chemicals) solution of varying polymer and cross-linker concentrations in n-heptane was spin coated onto quartz or silicon substrates and thermally cured in gentle vacuum to form elastomeric films with shear moduli of 0.1 MPa to 1 MPa and 200 nm to 10 μm thickness. Thickness and shear moduli were determined by ellipsometry (Nanofilm) and Bohlin rheometer or dynamic mechanical analyzer, respectively. All the films used were predominantly elastic with a nearly frequency independent value of $G'$ (storage modulus) exceeding $G''$ (loss modulus) by at least an order of magnitude. A flat stamp of silanized silicon wafer or glass was then brought into contact with the elastomer surface. For thicker (> 1 μm) films, the film was mounted on the objective of an optical microscope (Leica) to enable *in-situ* observations. The pattern lengthscale and the fractional area of contact between the stamp and the film surface, α, were characterized during different phases of adhesion and debonding. The elastic contact patterns formed relaxed to the flat film upon removal of stamp and the pattern at any fixed stamp position remained unaltered with time even after several hours. Further, neither several cycles of adhesion and debonding, nor the rate of stamp movement (lasting 4 minutes to 60 minutes to complete a cycle) produced any changes in the morphologies of the patterns and their lengthscales at a given *α*. For sub-micron films, the resulting structures were scanned with an AFM (Molecular Imaging) after the pattern was preserved by the UV-ozone induced surface hardening [10] of the silicon containing elastomer. This technique allowed permanent patterning and examination of patterns after removal of stamp, especially for thin (< 500 nm) films where the pattern had to imaged by AFM scanning. Figure 1 summarizes nearly bi-continuous labyrinths formed by gentle pressing of different thickness films against flat stamps. The structure consists of long ridges (contact zones) and intervening cavities (non-contact zones). Symbols in figure 2A represent the lengthscales of a large number of these labyrinth patterns, extracted from their FFTs, for different film thickness, shear modulus (0.01 -1 MPa) **a**nd the nature of stamp surface (silanized and non-silanized quartz and silicon wafer). For films in excess of about 1 micron thickness, the lengthscale of the pattern is indeed close **to** $n H$, with $n \sim 3$, independent of the shear modulus and the adhesive properties of the stamp. This is in agreement with the linear analysis [8] and 2-D simulations [9] for the onset of instability. However, there is a nonlinear increase in the prefactor *n* for sub-micron films, which, as we show below, is because of the increased importance of surface energy in thin films.

The previous experiments on contact instability have reported only the labyrinth structures [2] with a plane rigid stamp and its modified form of 2-D fingers with a curved flexible stamp. These labyrinth structures discussed above actually turn out to be an intermediate stage of morphological evolution during the processes of complete bonding (fig. 3; A1-A4) and complete detachment (fig. 3; D1-D4). Fig. 3, A1-A4 show the evolution of the instability as the stamp comes in increasing contact with the film in the process of bonding. The onset of instability occurs by the formation of discrete circular cross-section elastomeric pillars (A1) in contact with the stamp, which elongate and coalesce on further approach (A2) to make a labyrinth structure (A3). This almost bi-continuous intermediate structure finally morphologically phase inverts into isolated cavities (A4) that become increasingly circular on further approach. Fig. 3, D1-D4 summarize the morphological regimes of the stamp retraction or debonding phase where cavities grow (D2), elongate and coalesce (D3) and finally revert to pillars (D4) before complete detachment. These results provide insights into the process of contact cavitation, its origin, persistence and its influence on debonding at soft interfaces.

Interestingly, as shown in fig. 2B, the ratio $\lambda/H$ is nearly independent of the precise morphological phase during the entire debonding cycle, except close to complete detachment (fractional contact area < 0.1), where pillars begin to unbind.

The above experimental observations of three distinct morphologies (pillars, labyrinths and cavities) during adhesion and debonding, invariance of their lengthscales and the role of surface tension, can be explained by minimizing the total energy of the elastic film-stamp system. The total energy of the deformed films is composed of the stored elastic energy, the adhesive energy and the surface energy, given respectively by [8,9]:

$$U_T = \int_V W(\varepsilon)dV + \int_S U(d)dS + \int_S \gamma \sqrt{1+\left(\frac{\partial u_z}{\partial x}\right)^2 + \left(\frac{\partial u_z}{\partial y}\right)^2}\, dS \quad (1)$$

where $W(\varepsilon) = (\mu/2)\, \varepsilon : \varepsilon$ is the strain energy density, $\mu$ is the shear modulus of the incompressible film, $\varepsilon$ is the strain tensor in the film, $\gamma$ is the film surface energy, $U$ is the inter-surface interaction potential per unit area made up of adhesive interactions such as the van der Waals and $d(x,y)$ is the local inter-surface gap distance. The above energies can be obtained [9] from the displacement vector, $\boldsymbol{u}(x,y,z)$, the components of which are written as Fourier series. For example, the displacement of the free surface, $z = H$, in the normal direction is represented by [9]:

$$u_z(x,y) = \sum_{n_x=0}^{2Nx-1} \sum_{n_y=0}^{2Ny-1} v_{cc} \cos(k_{n_x} x)\cos(k_{n_y} y) + v_{cs} \cos(k_{n_x} x)\sin(k_{n_y} y) \\ + v_{sc} \sin(k_{n_x} x)\cos(k_{n_y} y) + v_{ss} \sin(k_{n_x} x)\sin(k_{n_y} y) \quad (2)$$

Where $v_{cc}$, $v_{cs}$, $v_{sc}$, $v_{ss}$ are the Fourier amplitudes and $k_n = \sqrt{k_{n_x}^2 + k_{n_y}^2}$ is the wave number. The boundary conditions include no-slip at the film-substrate interface and zero shear stress at the film surface [8,9].

Considering a single Fourier mode ($k$) and minimizing the total energy gives a relation for the onset of instability, where $Y(d)$ is the stiffness of the adhesive energy, $Y = \partial^2 U / \partial d^2 < 0$:

$$2kH \frac{(1+\cosh(2kH)+2k^2H^2)}{(\sinh(2kH)-2kH)} + \frac{\gamma}{\mu H}(kH)^2 = \frac{-HY}{\mu} \quad (3)$$

The onset of surface instability during adhesion occurs at a critical (minimum) attractive force, $Y$ corresponding to a critical wavenumber, $k$, for which the left hand side of eq. (3) is minimum with respect to $k$ (= $2\pi/\lambda$). With the inclusion of surface energy, the bifurcation mode in general has to be obtained numerically from eq. (3), which simplifies to the known linear stability results, $YH/\mu = 6.2$ and $kH = 2.12$ or $\lambda \sim 3H$ in the absence of surface energy contribution ($\gamma = 0$) [8,9]. It is interesting to note that the lengthscale of pattern at the onset of instability is predicted to be independent of the details of the adhesive interactions even if surface energy is significant. However, unlike the case of $\gamma = 0$, it now depends on nondimensional surface tension, $\gamma/\mu H$, the influence of which becomes increasingly important for softer ($\mu < 1$ MPa) and thinner films ($< 1$ μm). Inclusion of surface energy always increases the wavelength of instability because of its stabilizing influence in the form of an additional energy penalty which increases with a decrease in the wavelength. Indeed, the theoretical predictions from eq. (3) displayed in fig. 2A match the observed systematic increase in the wavelength as films get thinner than about 1 micrometer. Interestingly, a completely different scaling is obtained for very thin and soft films ($\gamma/\mu H \gg 1$), where the instability becomes long-wave ($kH \ll 1$). Eq. (3) now simplifies to: $-Y \approx (3\mu/H^3 \, k^2) + \gamma \, k^2$, which produces the energy minimizing wavelength to be: $\lambda/H \approx 2\pi \, (\gamma/3\mu H)^{1/4}$. The physics of very thin films is thus essentially different from the known short-wave scaling for thick films [2,3,8,9], where $\lambda/H \approx 3$ for $\gamma/\mu H \approx 0$.

Beyond the onset of instability, the energy minimization simulations were performed by a conjugate gradient scheme (which finds the local minimum closest to the initial configuration) to find the Fourier coefficients of the displacement fields that result in a minimum energy pattern for a given separation distance. The computations were performed for a full cycle of adhesion and debonding starting from the onset of instability. The surface interaction potential in simulations consisted of an attractive van der Waals component along with a short range Born repulsion [2,8,9], $U(d) = -A/12\pi d^2 + B/d^8$, where A is the Hamaker constant ($A = 10^{-20}$ J), B is chosen to give a realistic energy of adhesion at contact, $\sim 30$ mJ/m$^2$ [11] and $d(x,y)$ is the local inter-surface distance. Details of the method, which was implemented earlier in the 2-D geometry, can be found elsewhere [9].

Results of 3-D simulations for the bonding (A1-A4) and debonding (D1-D4) phases are displayed in fig. 4, which indeed parallel the morphological phases and their transitions observed in the corresponding experiments (fig. 3). Interestingly, based on a large number of simulations, it could be concluded that regardless of the film thickness, shear modulus and details of the adhesive potential, the development of morphological features is very similar in the same stages or regimes of bonding and debonding as characterized by the fractional area of contact with the stamp.

The critical length scale of the linear theory at the onset of bifurcation is not in general expected to predict the length scales of the nonlinear, morphologically distinct structures during different stages of approach and debonding. This aspect is addressed by our experiments and 3-D

8simulations as summarized in fig. 2B. The interesting feature uncovered by experiments and supported by the simulations is that the lengthscale of different morphological phases is in fact nearly invariant regardless of the precise morphology, except very close to complete debonding where columns begin to detach (when α = fractional contact area < 10%). The persistence of a unique length scale during bonding and debonding cycle is because at the onset of instability, a partial adhesive contact is established with the stamp, leading to a metastable structure pinned in a local minimum of the total energy. During further approach or retraction of the stamp, this metastable state surrounded by energy barriers is merely modified by an enlargement or reduction in its contact area, rather than by the formation of new structures of different wavelengths, even though these structures may be energetically more favorable. The simulations reported indeed capture this aspect because the conjugate gradient search employed for energy minimization tracks the nearest energy minimum closet to the previous configuration, starting from the onset of instability.

In summary, we have shown, both experimentally and theoretically, that the cycle of adhesion and debonding to a soft elastic surface manifests three distinct morphological phases (pillars, labyrinths and cavities), but the length scale of the structures is rather robust against these morphological transitions. The non-dimensional wavelength, $\lambda/H$ depends only the non-dimensional surface tension, $(\gamma/\mu H)$. For relatively thick films (> 1 micrometer), the linear theory scaling, $\lambda \sim 2.96\, H$ is respected, but for thinner films, increased surface to volume ratio increases the wavelength, which now depends nonlinearly on the film thickness. This is of special interest in polymeric opto-electronic coatings and soft lithography where films are getting progressively thinner [1,5-7]. In addition to elucidating the underlying physics and anatomy of soft adhesion, these results have the potential to develop into a new soft lithography technique [12] which utilizes the self-organized elastic deformations engendered by pre-patterned stamps. Such a technique can provide unprecedented flexibility for *in-situ* morphological modulation (by approach and retraction of the stamp) and length scale control (by film thickness).

**Acknowledgements**

We thank M. K. Chaudhury and A. Ghatak for illuminating discussions. This work was supported by the DST Unit on Nanosciences at IIT Kanpur.

[*]ashutos@iitk.ac.in

**Figures**

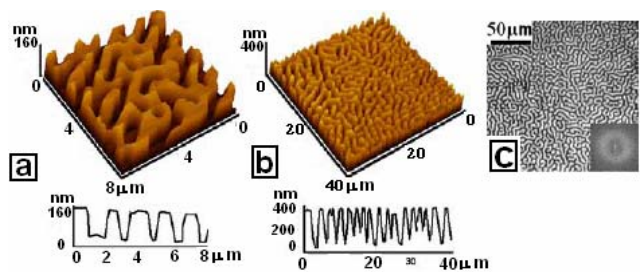

Fig. 1. Samples of labyrinth patterns formed in thin films by gentle pressing of a flat rigid contactor: (a) H = 280 nm, λ = 1.58 ± 0.15 μm, (b) H = 490 nm, λ = 2.33 ± 0.29 μm, (c) H = 1.68 μm, λ = 5.33 ± 0.27 μm. Insets are the corresponding FFT images. (a), (b) μ ~ 0.5 MPa; (c) μ ~1 MPa

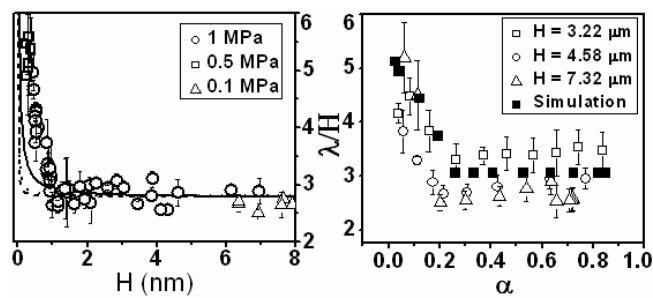

Fig. 2: (A) Wavelengths of the labyrinth patterns of the type shown in fig. 1; μ = 0.1- 1 MPa. Lines represent the linear stability analysis for μ = 0.1 MPa and 1 MPa and γ = 22 mJ/m$^2$. (B) Variation of the wavelength with the fractional contact area during debonding; μ= 1 MPa.

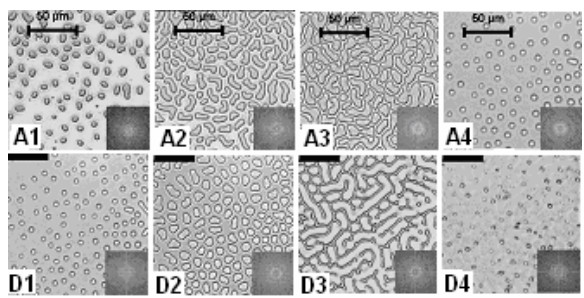

Fig 3: Morphological evolution during adhesion in the bonding phase (A1- A4) and debonding phase (D1-D4) of the stamp to a 3.2 μm thick film; μ ~ 1 MPa. (A1) isolated columns at fractional contact area, α ~ 0.25, (A2) elongated columns and coalescence, α ~ 0.51, (A3) labyrinths, α ~ 0.68, and (A4) isolated cavities, α ~ 0.76. (D1) isolated cavities, α ~ 0.74, (D2) expanded cavities approaching coalescence, α ~ 0.56, (D3) labyrinths, α ~ 0.37, and (D4) isolated columns approaching complete detachment, α ~ 0.03. Scale bar = 50 μm.

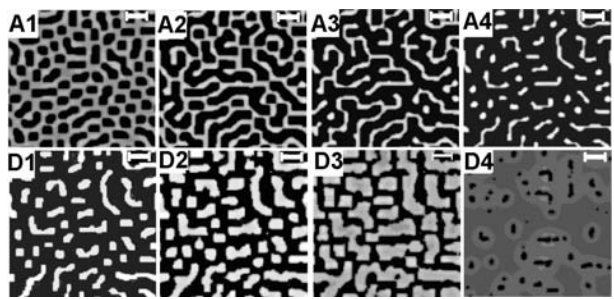

Fig 4: Simulation of morphological evolution during the bonding phase (A1-A4) and debonding phase (D1-D4) of the film shown in fig. 3. Darker regions are in contact and brighter regions are cavities. The fractional contact areas in A1 to A4 are 0.44, 0.51, 0.68 and 0.81, respectively, and in the images D1 to D4 are 0.74, 0.56, 0.37 and 0.03, respectively. Scale bar = 12.8 μm. The simulations also mirror the cellular nature of experimental structures of fig. 4.